\documentclass[11pt]{article}
\usepackage{amssymb}
\def\ds{\displaystyle}

\makeatletter
\@addtoreset{equation}{section}

\newtheorem{theorem}{Theorem}[section]
\newtheorem{examp}{Example}[section]
\newtheorem{coroll}{Corollary}[section]
\newtheorem{examps}{Examples}[section]

\newtheorem{lemma}{Lemma}[section]
\newtheorem{remark}{Remark}[section]

\newtheorem{remarks}[remark]{Remarks}
\newtheorem{proposition}{Proposition}[section]
\newtheorem{definition}{Definition}[section]

\def\le{\left}
\def\m{\mathop}
 \def\tr{{\rm Tr}}

\def\ri{\right}
\def\br{\begin{remark}\rm\small}
\def\1{{\bf 1}}
\def\er{\end{remark}}
\def\bt{\begin{theorem}\rm}
\def\et{\end{theorem}}
\def\bc{\begin{coroll}\rm}
\def\ec{\end{coroll}}
\def\brs{\begin{remarks}.\\ \rm\small\begin{enumerate}}
\def\ers{\end{enumerate}\end{remarks}}
\def\bx{\begin{examp}\small}
\def\ex{\end{examp}}
\def\bl{\begin{lemma}\small}
\def\el{\end{lemma}}
\def\bxs{\begin{examps}. \rm\begin{enumerate}}
\def\exs{\end{enumerate}\end{examps}}
\def\bd{\begin{definition}}
\def\ed{\end{definition}}
\def\bp{\begin{proposition}\rm}
\def\ep{\end{proposition}}
\def\be{\begin{equation}}
\def\ee{\end{equation}}

\def\bea{\begin{eqnarray}}
\def\eea{\end{eqnarray}}
\def\beas{\begin{eqnarray*}}
\def\eeas{\end{eqnarray*}}

\def\C{{\mathbb C}}

\def\R{{\mathbb R}}
\def\N{{\mathbb N}}

\begin{document}
\begin{flushright}
CRM-2916 (2003)\\
\hfill Saclay-T03/028
\end{flushright}
\vspace{0.2cm}
\begin{center}
\begin{Large}
\textbf{Mixed Correlation Functions of  the Two-Matrix Model}
\end{Large}\\
\vspace{1.0cm}
\begin{large} {M.
Bertola}$^{\dagger\ddagger}$\footnote{bertola@mathstat.concordia.ca}, 
 { B. Eynard}$^{\ddagger
\star}$\footnote{eynard@spht.saclay.cea.fr}
\end{large}
\\
\bigskip
\begin{small}
$^{\dagger}$ {\em Department of Mathematics and
Statistics, Concordia University\\ 7141 Sherbrooke W., Montr\'eal, Qu\'ebec,
Canada H4B 1R6} \\ 
\smallskip
$^{\ddagger}$ {\em Centre de recherches math\'ematiques,
Universit\'e de Montr\'eal\\ C.~P.~6128, succ. centre ville, Montr\'eal,
Qu\'ebec, Canada H3C 3J7} \\
\smallskip
$^{\star}$ {\em Service de Physique Th\'eorique, CEA/Saclay \\ Orme des
Merisiers F-91191 Gif-sur-Yvette Cedex, FRANCE } \\
\end{small}
\bigskip
\bigskip
%%%%%%%%%%%%%%%%%%%%%%%%%%%%%%%%%%  Abstract %%%%%%%%%%%%%%%%%%%%%%%%%%%%%%%
{\bf Abstract}
\end{center}
\begin{center}
\begin{small}
\parbox{13cm}{
We compute the correlation functions mixing  the powers of 
two non-commuting random matrices within the  same trace. 
The angular part of the integration was partially known in the literature
\cite{morozov,shata}: we  pursue the calculation and carry out the
eigenvalue integration reducing the problem to the construction of
the associated biorthogonal polynomials.
The generating function of these correlations becomes then
 a determinant involving the recursion coefficients
of the biorthogonal polynomials.
}
\end{small}
\end{center}
%%%%%%%%%%%%%%%%%%%%%%%%%%%%%%%%%%%%%%%%%%%%%%%%%%%%%%%%%%%%%%%%%%%%%%%%%%%%%%%%
%
%                                                                               
%
%                       Introduction begins...                                   
%
%                                                                               
%
%                                                                               
%
%%%%%%%%%%%%%%%%%%%%%%%%%%%%%%%%%%%%%%%%%%%%%%%%%%%%%%%%%%%%%%%%%%%%%%%%%%%%%%%%
%
\
\section{Introduction and main result}

Random matrix models were first introduced in the context of nuclear physics in 
order to describe the energy levels statistics for very large
nuclei.
Wigner proposed the hypothesis that these were distributed as the eigenvalues of a 
matrix with random entries.
Later random matrix models were used in many areas of physics and mathematics \cite{DGZ,Guhr,Mehta}.

An important application of random matrices is to 2d gravity, that is, statistical physics on a random surface.
In fact, the perturbative expansion of a matrix integral can be
accomplished by drawing Feynman graphs on fixed-genus surfaces. 
 Matrix integration can therefore encode the summation over the  set of discretized surfaces 
(possibly carrying some type of matter).

When the parameters of the model are fine-tuned near a critical point the 
average graph's size diverges and macroscopic graphs dominate the sum,
so that 
 suitable critical limits can represent  
statistical models over smooth surfaces.
The  smooth surfaces that one describes with the aid of matrix models 
 have properties of scale invariance: this means that the  critical points of matrix 
integrals are related to representations of the conformal group. We
recall that its  finite dimensional representations are classified by 
two integers $(p,q)$:
it is known that one-matrix-models provide
instances of  $(p,2)$-irreps only, where as 
two-matrix-models allow to obtain $(p,q)$--representations.

Possibly the first such application of the two--matrix--model was to describe the Ising model on a 
random surface; in this case the Ising ferromagnetic transition corresponds to the conformal 
minimal model $(3,4)$.
To see this, one should associate a color (or spin) $+$ or $-$ to each matrix
so that the
vertices of one matrix are labelled with a plus sign
 and the vertices of 
the other matrix with a minus sign.
 Then the Feynman graphs generated by a two 
matrix model represent discrete surfaces carrying spins (+ and -),
i.e. an  Ising model on a random surface.

The correlation functions of random matrices generate discrete surfaces with boundaries, 
and thus are in relationship with boundary conformal theory.
A formula for correlation functions representing surfaces with mono-colored boundaries 
has been known since \cite{eynardmehta}.
It is the aim of this paper to give a formula for a mixed correlation function,
i.e. the generating function for discrete surfaces with a bi-colored boundary.

The 2-matrix model has attracted a lot of attention recently,
 and important progress have been made in the study of the associated bi-orthogonal 
 polynomials \cite{BHE1, BHE3, kapaev}.
Here, we will express the mixed correlation function in terms of the bi-orthogonal polynomials.

\subsection{Definition and notation}
We consider two $N\times N$ Hermitian matrices $M_1,M_2$,
with a probability measure
\bea
&& {\rm d}\mu(M_1,M_2) := {\mathcal Z_N}^{-1}{\rm d}M_1{\rm d}M_2
\exp\bigg[-\tr(V_1(M_1)+V_2(M_2) -M_1M_2)\bigg]\ , \nonumber \\
&& {\mathcal Z_N}:= \int\int{\rm d}M_1{\rm d}M_2
\exp\bigg[-\tr(V_1(M_1)+V_2(M_2) -M_1M_2)\bigg]\label{measure}\ ,
\eea
where ${\rm d}M_1{\rm d}M_2$ is the product of Lebesgue measures of
all the independent real and imaginary parts of the components of the
two matrices divided by the square of the volume of the unitary group
$U(N)$ (for later convenience). The functions 
$V_1$ and $V_2$ are called the potentials and  must be chosen so as to
make  the integral convergent. The normalization factor
${\mathcal Z_N}$ is called  the ``partition function'', where the name
``function'' refers to its dependence on the two potentials. 

One can rewrite the measure in term of eigenvalues and angular integrals 
\cite{IZ}:
\bea
&& {\rm d}\mu(M_1,M_2) := {\mathcal Z_N}^{-1} \Delta(X)^2\Delta(Y)^2
\exp\bigg[-\sum_{i=1}^N(V_1(x_i)+V_2(y_i) )\bigg] {\rm e}^{U^\dagger X
  U V^\dagger Y V}{\rm d}U{\rm d}V\prod_{i=1}^N{\rm d}x_i{\rm
  d}y_i\ , \nonumber \\
&& X:={\rm diag}(x_1,\dots,x_N),\ Y:={\rm diag}(y_1,\dots,y_N),\ U,V\in 
U(N)\nonumber
\\
&& {\mathcal Z_N}:= \int\int\prod_{i=1}^N{\rm d}x_i{\rm
  d}y_i \Delta(X)^2\Delta(Y)^2{\rm d}U{\rm d}V
\exp\bigg[-\sum_{i=1}^N(V_1(x_i)+V_2(y_i) )\bigg] {\rm e}^{U^\dagger X
  U V^\dagger Y V}\label{measure2}\ .
\eea
where ${\rm d}U{\rm d}V$ is the product of the {\em normalized}  Haar measure over  $U(N)\times U(N)$.
In the original two Hermitian matrix model, the integration path for the $x_i$'s 
and $y_j$'s is the real axis (location of the eigenvalues of a Hermitian 
matrix), however, the model can be generalized
to include complex paths or their homology classes in case  the potentials
are holomorphic or meromorphic \cite{Berto,BHE1}.

\subsection{Correlation functions}

In applications of the two--matrix--model to statistical physics on a random 
surface, one is interested in computing correlation functions
involving traces of products of powers of $M_1$ and $M_2$.
Each such correlation function can be expanded in Feynman graphs, which 
represent discrete surfaces with boundaries:
the number of boundaries is the number of traces,
the length of each boundary is the total power of $M_1$ plus the total power of 
$M_2$ within the trace \cite{DGZ}.

For instance:
$
\left< \tr M_1^r \right>
$
is the generating function for discrete surfaces with one boundary (a circle)
 of length $r$ made of $+$ spins only.
$
\left< \tr M_2^s \right>
$
is the generating function for discrete surfaces with one boundary of 
length $s$ made of $-$ spins only.
$
\left< \tr M_1^r M_2^s \right>
$
is the generating function for discrete surfaces with one bi-colored 
boundary of length $r+s$ made of $r$ $+$ spins, followed by $s$ $-$ spins.

More generally, 
$
\left< \tr M_1^{r_1} M_2^{s_1} M_1^{r_2} M_2^{s_2} \dots M_1^{r_n} M_2^{s_n} 
\right>
$
is the generating function for discrete surfaces with one $2n$-colored boundary
of length 
$\sum_i r_i + \sum_i s_i$ made of $r_1$ $+$ spins, 
followed by $s_1$ $-$ spins, followed by $r_2$ $+$ spins,~$\dots$, 
followed by $s_n$ $-$ spins.

One may also be interested in ``multi-loop'' correlators (i.e. more than one 
boundary), for instance:
$
\left< \tr M_1^{r_1}  \tr M_1^{r_2} \right>_{\rm conn}
$
is the generating function for discrete surfaces with two spin $+$ boundaries,
one of length $r_1$, the other of length $r_2$.
More generally, one may consider correlation functions involving an arbitrary 
number of traces, each containing arbitrary words of $M_1$ and $M_2$.

The correlation functions, with an arbitrary number of traces, with each trace 
containing powers of only one matrix have been known since the work of 
\cite{eynardmehta}.
They can be expressed in terms of bi-orthogonal polynomials.

The aim of the present article is to express the mixed correlation function
\be
\left< \tr M_1^r M_2^s \right>
\ee
in terms of bi-orthogonal polynomials too, and confirm that the key property of 
these models is that all relevant spectral
statistics can be reduced to the computation of the corresponding
biorthogonal polynomials \cite{BHE2,eynardmehta,mehta}.

\subsection{Bi-orthogonal polynomials}
 Two sequences of monic polynomials
\be
\pi_n(x) = x^n + \cdots , \qquad \sigma_n(y)=y^n + \cdots, \qquad n=0,1,\dots
\ee
are called biorthogonal if they are ``orthogonal''  with respect to a coupled measure on the product space:
\be
\int_\R\!\!\int_\R\!\!\! {\rm d}x\, {\rm d}y \,\, \pi_n(x)\sigma_m(y) {\rm
e}^{- V_1(x)- V_2(y) +xy} = h_n\delta_{mn} ,\qquad h_n\neq 0\ \forall n\in\N\label{norms}
\ee 
where $V_1(x)$ and $V_2(y)$ are  the functions (called {\em potentials})
appearing in the two-matrix model measure (\ref{measure}).
It is convenient to introduce the associated quasipolynomial
differentials defined by the formulas
\bea
&&\psi_n(x):= \frac 1{\sqrt{h_{n-1}}}\pi_{n-1}(x){\rm e}^{-V_1(x)}{\rm d}x\\
&&\phi_n(y):= \frac 1{\sqrt{h_{n-1}}}\sigma_{n-1}(y){\rm e}^{-V_2(y)}{\rm
d}y\ .\label{quasidifferentials}
\eea
In terms of these two sequences of differentials the multiplications
by $x$ and $y$ respectively are represented by semiinfinite square
matrices $Q = [Q_{ij}]_{i,j\in \mathbb N^*}$ and  $P = [P_{ij}]_{i,j\in
  \mathbb N^*}$ according to the formulae
\bea
&&x\psi_n(x) = \sum_m Q_{n,m}\psi_m(x)\ ;\ \ \ y\phi_n(y) = \sum_m
P_{m,n}\phi_m(y) \nonumber \\
&&Q_{n,m}=0=P_{m,n}, \ {\rm if}\ n>m+1.\label{mulxy}
\eea
The matrices $P$ and $Q$ have a rich structure and satisfy the
``string equation'' $[P,Q]=id$. However we do not need any of their
properties except for eq. (\ref{mulxy}) to derive our present results,
and therefore we refer for further details to \cite{Berto,BHE1,BHE2,BHE3} where
these models are studied especially in the case of polynomial potentials. We 
also point out that the model can easily be generalized to accommodate
contours of integration other than the real axes \cite{Berto,BHE1} leaving
intact all the properties which are relevant to the following computations.

\subsection{The main result}
Our goal is to prove a  formula for the generating function of the
correlators
\bea
\langle\tr({M_1}^r{M_2}^s)\rangle_{V_1,V_2} :=  {1\over {\mathcal Z_N}}\int  
 {\rm d} M_1{\rm d}M_2
 \tr({M_1}^r{M_2}^s)\exp\le(-\tr\le(V_1(M_1)+V_2(M_2)-M_1M_2\ri)\ri)\ .
\label{corrfunc}
\eea
By generating function we mean the formal double Laurent series 
\bea
\le\langle\tr\le(\frac 1{x-M_1}\frac
1{y-M_2}\ri)\ri\rangle_{V_1,V_2}:= \sum_{r,s} x^{-r-1}y^{-s-1}
\langle\tr({M_1}^r{M_2}^s)\rangle_{V_1,V_2}.
\label{genfunc}
\eea
The main obstacle to this sort of computations so far was posed by
the  ``angular integration'' over the unitary group $U(N)$.

One can trace in the literature various attempts at this computation using the 
loop equations \cite{eynardchain,eynard,staudacher}.
A closed formula was found in the large $N$ limit
 in \cite{eynard}, but an exact formula for finite $N$ was never 
derived.

Our strategy is that of reducing the computation of (\ref{corrfunc})
or even better (\ref{genfunc}) to the computation of the corresponding
biorthogonal polynomials associated with the measure (\ref{measure}).

We can now write down the  main  result of the paper which is both simple and 
beautiful:
\bea
\hbox{
\framebox{\parbox{12cm}{
\begin{center}
$\ds
\le\langle\tr\le(\frac 1{x-M_1}\frac
1{y-M_2}\ri)\ri\rangle_{V_1,V_2} = 1- \det\le[\1_{N} -\m{\pi}_N\frac
  1{x-Q}\frac 1{y-P}\m{\pi}_N\!^t \ri] $
\end{center}
}}}\label{beauty}
\eea
where $P$ and $Q$ are the matrices in eq. (\ref{mulxy}) and  $\ds \m{\pi}_N$ 
denotes the projector $\C^\infty\mapsto \C^N$  onto the span of
the first $N$ canonical basis vectors, i.e., the $N\times \infty$
matrix with nonzero entries $\ds\m{\pi}_N\!_{i,i}=1,\ i=1,\dots N$.
Formula (\ref{beauty}) should be properly understood in the sense of
an identity of formal Laurent  series in the variables $x$ and $y$,
although it would be possible to give an analytic meaning to both sides.
For example from (\ref{beauty})  one can easily obtain the
following identities
\bea
\le\langle \tr({M_1}^r{M_2})\ri\rangle_{V_1,V_2} = \tr(Q^r P\Pi ) -\frac 1 2
\sum_{j=0}^{r-1} \le(\tr(Q^j\Pi)\tr(Q^{r-1-j}\Pi) - 
\tr(Q^j\Pi Q^{r-1-j}\Pi)\ri)
\eea
\bea
\le\langle \tr({M_1}^r{M_2}^2)\ri\rangle_{V_1,V_2} &=& \tr( Q^rP^2
\Pi) -\!\!\!\!\!\! \sum_{k_1+k_2=r-1} \!\!\!
\le(  \tr(Q^{k_1}P\Pi) \tr(Q^{k_2}\Pi)
- \tr(Q^{k_1}P\Pi Q^{k_2}\Pi) 
\ri)\\ 
&&\hspace{-120pt}+ \hspace{-22pt} \sum_{k_1+k_2+k_3=r-2}\!\!\!
 \le(
\frac 1 3\tr(\Pi Q^{k_1}\Pi Q^{k_2} \Pi Q^{k_3})  
\!-\!\frac 1 2 \tr(\Pi Q^{k_1}\Pi Q^{k_2}) \tr(\Pi Q^{k_3})
\!+\! \frac 1 6\tr(\Pi Q^{k_1})\tr(\Pi Q^{k_2})\tr( \Pi Q^{k_3})  
\ri)
\eea
%\be
%\le\langle\tr(M_1^2M_2^2)\ri\rangle_{V_1,V_2} = \tr(P^2\Pi Q^2) - (N-1)\tr(P\Pi
%Q) - \tr(P\Pi)\tr(Q\Pi) + \tr(P\Pi Q\Pi) + \le(N\atop 3 \ri)
%\ee
where now $\Pi = \pi^t\pi$ is  the semiinfinite square
matrix with nonzero entries $\Pi_{i,i}=1,\ i=1,\dots N$.

In fact in the course of the proof of eq. (\ref{beauty}) we will also
prove the following strong result for the correlations of 
$\langle \tr({M_1}^r {M_2}^s) \rangle_{V_1,V_2}$ which naturally
extends to arbitrary (analytic) functions $f(x),g(y)$:
\bea
&& \hspace{-1cm}\langle \tr({M_1}^r{M_2}^r) \rangle_{V_1,V_2} = \\
&& \hspace{-1cm}=\sum_{n=0}^{\min(r,s)}
\frac{(-1)^n}{(n+1)!} \int\cdots\int
{\rm e}^{\sum_k x_ky_k} \frac { 
\det\pmatrix{
1&\cdots&1\cr
x_{1}& \cdots& x_{{n+1}}\cr
\vdots& &\vdots\cr
x_{1}^{n-1}& \cdots & x_{{n+1}}^{n-1}\cr
x_{1}^r & \cdots & x_{n+1}^r }}{
\det\pmatrix{
1&\cdots&1\cr
x_{1}& \cdots& x_{{n+1}}\cr
\vdots& &\vdots\cr
x_{1}^{n}& \cdots & x_{{n+1}}^{n}}} 
\frac { 
\det\pmatrix{
1&\cdots&1\cr
y_{1}& \cdots& y_{{n+1}}\cr
\vdots& &\vdots\cr
y_{1}^{n-1}& \cdots & y_{n+1}^{n-1}\cr
y_{1}^s & \cdots & y_{n+1}^s }}{
\det\pmatrix{
1&\cdots&1\cr
y_{1}& \cdots& y_{n+1}\cr
\vdots& &\vdots\cr
y_{1}^{n}& \cdots & y_{n+1}^n }} \det\bigg[K_{12}(x_k,y_\ell)\bigg]_{k,\ell\leq n+1}
\label{RScorr}\ .
\eea
The formula can be extended to arbitrary analytic functions $f(x)$ and
$g(y)$ to give 
\bea
&& \hspace{-1cm}\langle \tr(f(M_1)g(M_2) ) \rangle_{V_1,V_2} = \\
&& \hspace{-1cm} =\sum_{n=0}^{\infty}
\frac{(-1)^n}{(n+1)!} \int\cdots\int{\rm e}^{\sum_k x_ky_k} 
 \frac { 
\det\pmatrix{
1&\!\!\!\!\!\!\cdots&1\cr
x_{1} &\!\!\!\!\!\!\cdots&\!\!\!\!\!\! x_{{n+1}}\cr
\vdots &&\vdots\cr
x_{1}^{n-1}&\!\!\!\!\!\! \cdots &\!\!\!\!\!\! x_{{n+1}}^{n-1}\cr
f(x_{1})  &\!\!\!\!\!\!\cdots  &\!\!\! \!\!\!f(x_{n+1}) }}{
\det\pmatrix{
1&\cdots&1\cr
x_{1}& \cdots& x_{{n+1}}\cr
\vdots& &\vdots\cr
x_{1}^{n}& \cdots & x_{{n+1}}^{n}}} 
\frac { 
\det\pmatrix{
1&\!\!\!\!\!\!\cdots&1\cr
y_{1}&\!\!\!\!\!\! \cdots&\!\!\!\!\!\! y_{{n+1}}\cr
\vdots& &\vdots\cr
y_{1}^{n-1}&\!\!\!\!\!\! \cdots &\!\!\!\!\!\! y_{n+1}^{n-1}\cr
g(y_{1}) &\!\!\!\!\!\! \cdots &\!\!\!\!\!\!g( y_{n+1}) }}{
\det\pmatrix{
1&\cdots&1\cr
y_{1}& \cdots& y_{n+1}\cr
\vdots& &\vdots\cr
y_{1}^{n}& \cdots & y_{n+1}^n }} \det\bigg[K_{12}(x_k,y_\ell)\bigg]_{k,\ell\leq n+1}
\label{fgcorr}
\eea
 Here we have used the kernel constructed from the quasipolynomials \cite{eynardmehta}, 
\be
K_{12}(x,y):= \sum_{n=1}^{N}\psi_n(x)\phi_n(y) = \sum_{n=0}^{N-1}
 \frac {\pi_n(x)\sigma_n(y){\rm e}^{-V_1(x)-V_2(y)}}{h_n} {\rm d}x\,
 {\rm d}y\ .  \label{kernel}
\ee
 
%%%%%%%%%%%%%%%%%%%%%%%%%%%%%%%%%%%%%%%%%%%%%%%%%%%%%%%%%%%%%%%%%%%%%%%%%%%%%%%%
%
%                                                                               
%
%                               End of Introduction                             
%
%                                                                               
%
%                                                                               
%
%%%%%%%%%%%%%%%%%%%%%%%%%%%%%%%%%%%%%%%%%%%%%%%%%%%%%%%%%%%%%%%%%%%%%%%%%%%%%%%%
%

\section{Proofs of formul\ae\ (\ref{beauty}) and (\ref{RScorr})}
%
%We consider the two--matrix model with polynomial potentials
%\be
%{\rm d}\mu (M_1,M_2):= {\rm d} M_1{\rm d}M_2
%\exp\le(-\tr\le(V_1(M_1)+V_2(M_2)-M_1M_2\ri)\ri),
%\ee
%where the integration is taken over the set of pairs of $N\times N$
%{\em normal} matrices $M_1,M_2$, with  spectrum on certain
%curves on the complex plane.
%We denote by ${\mathcal Z}_N$ the partition function and
We want to 
compute the expectation values
\be
\le\langle\tr({M_1}^r{M_2}^s)
\ri\rangle_{V_1,V_2}:= \frac 1{{\mathcal Z}_N} \int  {\rm d}\mu (M_1,M_2)
\tr({M_1}^r{M_2}^s).
\label{eq1}
\ee
 We denote by
$\{x_i\}_{i=1..N}$ and $\{y_i\}_{i=1..N}$ the spectra of the two
matrices $M_1,\ M_2$ and by $U\in U(N)$ the relative angles.
 Then we can reduce the integral (\ref{eq1}) to an integral over the spectra of $M_1, M_2$ and
 the unitary group of the relative angles. Indeed we have
\bea
&& \langle\tr(M_1^rM_2^s)\rangle_{V_1,V_2} := \frac 1{{\mathcal Z}_N} \int  {\rm d}\mu (M_1,M_2)
\tr(M_1^rM_2^s) = \\
&& =\frac 1{{\mathcal Z}_N} \int \prod_{i=1}^N {\rm d}x_i{\rm d}y_i\,\, {\rm
e}^{-V_1(x_i)-V_2(y_i)} \Delta^2(X)\Delta^2(Y)\sum_{i,j=1}^N {x_i}^r{y_j}^s
\int_{U(N)}{\rm d}U  |U_{ji}|^2 {\rm
e}^{\tr(XU^\dagger YU)}\ ,\label{eq2}
\eea
where $X = {\rm diag}(x_1,\dots,x_N)$, $Y= {\rm diag}(y_1,\dots,y_N)$
and  $\Delta(X)$, $\Delta(Y)$ denote the Vandermonde determinants.
The computation requires the knowledge of the
two--points correlators for the unitary integral in
eq. (\ref{eq2}) which we analyze in the next subsection.

\subsection{Two--points correlator for the unitary integral}

The computation of this quantity has been considered in the finite $N$
regime in the two
papers \cite{morozov,shata}. In \cite{shata} was described a complete algorithm that allows  the 
construction of a formula for  the
most general correlator 
\be
\langle U_{i_1j_1}U^\dagger_{k_1l_1}\cdots
U_{i_nj_n}U^\dagger_{k_nl_n}\rangle_{U(N)} 
 := \int_{U(N)} \!\!\!{\rm d}U \,  U_{i_1j_1}U^\dagger_{k_1l_1}\cdots
U_{i_nj_n}U^\dagger_{k_nl_n} {\rm e}^{{\rm Tr} (XU^\dagger YU)}\ .
\ee
Such algorithm involves the introduction of  the parametrization of the
unitary group given by the Gel'fand-Tsetlin coordinates associated to one
of the two matrices $M_i$; however the
computation is highly involved and in \cite{shata} the problem was not carried 
through
to a completely manageable formula.\par
On the other hand in \cite{morozov} a very simple closed formula was
proposed for the two--point correlator $\langle|U_{ji}|^2\rangle_{U(N)}$ in
terms of a generating function, that is
\bea
&&\hspace{-1.6cm}\sum_{i,j}
a_ib_j\langle|U_{ji}|^2\rangle_{U(N)} = 
\sum_{i,j=1}^N {a_i}{b_j}
\int_{U(N)}{\rm d}U  |U_{ji}|^2 {\rm
e}^{\tr(XU^\dagger YU)} = \\
&& \hspace{-1.5cm}=\frac 1{\Delta(X)\Delta(Y)} 
\sum_{\rho\in S_N} \epsilon(\rho) {\rm e}^{\sum x_\ell y_{\rho(\ell)}}
\sum_{n=0}^{N-1} (-1)^{n}\!\!\!\!\!\!\!\!\!\!
\sum_{i_1<i_2<\dots<i_{n+1}}
\hspace{-20pt}
 \frac { 
\det\pmatrix{
1&\cdots&1\cr
x_{i_1}& \cdots& x_{i_{n+1}}\cr
\vdots& &\vdots\cr
x_{i_1}^{n-1}& \cdots & x_{i_{n+1}}^{n-1}\cr
a_{i_1} & \cdots & a_{i_{n+1}}}}{
\det\pmatrix{
1&\cdots&1\cr
x_{i_1}& \cdots& x_{i_{n+1}}\cr
\vdots& &\vdots\cr
x_{i_1}^{n}& \cdots & x_{i_{n+1}}^{n}}} 
\frac { 
\det\pmatrix{
1&\cdots&1\cr
y_{\rho(i_1)}& \cdots& y_{\rho(i_{n+1})}\cr
\vdots& &\vdots\cr
y_{\rho(i_1)}^{n-1}& \cdots & y_{\rho(i_{n+1})}^{n-1}\cr
b_{\rho(i_1)} & \cdots & b_{\rho(i_{n+1})}}}{
\det\pmatrix{
1&\cdots&1\cr
y_{\rho(i_1)}& \cdots& y_{\rho(i_{n+1})}\cr
\vdots& &\vdots\cr
y_{\rho(i_1)}^{n}& \cdots & y_{\rho(i_{n+1}) }^{n}}}\ ,\label{moroformula}
\eea
with the understanding  that the Haar measure of the unitary group $U(N)$
has been normalized to unity.
This formula will be the starting point of our analysis: however the
 author of \cite{morozov} did not actually prove the formula but just
 made an educated (and -as it turns out- correct) guess. 

Therefore, before proceeding to proving our main result (\ref{beauty})
we want to fill in
the gaps between the full but unpractical algorithm given in \cite{shata} and
the practical but unproven formula in \cite{morozov}.
We do not need the full generality of  \cite{shata}: our
departure  point is  formula (1.4) ibidem, restricted to the particular case of 
the two--point
correlator.
For the ease of the reader we rewrite the aforementioned  formula in the notation of
our present paper
\bea
&& \hspace{-20pt}\langle U_{1j_1}U^\dagger_{k_11}\cdots U_{1 j_n}U_{k_n1}^\dagger
\rangle_{U(N)} =  \frac
        {\delta_{j_1k_1}\cdots\delta_{j_nk_n}}{\Delta(X)\Delta(y_2,\dots,y_N)}
\le(\prod_{k=1}^{N-1} \int_{x_k}^{x_{k+1}}\!\!\!\!\!\!\!{\rm d}\xi_k\ri)
 \frac {\ds \prod_{\ell=1}^{N-1}(\xi_\ell-x_{j_1})
  \cdots \prod_{\ell=1}^{N-1} (\xi_\ell-x_{j_n})} {\ds \prod_{\ell\neq
    j_1} (x_{\ell}-x_{j_1})\cdots \prod_{\ell\neq
    j_n}(x_\ell-x_{j_n})} \times\nonumber  \\
&&  \times \exp\le[ \sum_{k=1}^N
  x_ky_1-\sum_{k=1}^{N-1} \xi_ky_1\ri] \det\le[{\rm
  e}^{\xi_iy_{j+1}}\ri]_{i,j=1..N-1} \label{1.4}\ .
\eea
Here the Haar measure of the unitary group has been normalized to unity, 
$\Delta(y_2,\dots,y_N)$ denotes the Vandermonde
determinant of the $N-1$ numbers $y_2,...y_N$, while $\Delta(X)$ is
the short form for the Vandermonde determinant of the whole spectrum
of the matrix $X$.\par
Eq. (\ref{1.4}) was proven  rigorously in \cite{shata} 
for $x_1<x_2<\dots <x_N$; however it is straightforward to realize that
the unitary integral in eq. (\ref{eq2}) defines an analytic (in fact
entire) function of the variables $X$ and $Y$. Therefore the result
extends to 
statistical ensembles of pairs $M_1,\ M_2$ of 
{\em normal matrices}\footnote{We recall that a  normal matrix is a matrix that
  commutes with its Hermitian-transposed. Any such matrix can be
  diagonalized using a unitary transformation (and vice-versa).}
 with their spectrum on arbitrary paths on the
complex plane. Moreover the  restriction on the order of the
spectrum can be lifted because the result is analytic in the variables
$x_1,\dots,x_N$ and can be analytically continued to $\C^N$.\par
 We now 
set $n=1$ in eq. (\ref{1.4}) and hence $k_1=j_1=i$. Then
we have :
\bea
&& \langle |U_{1i}|^2 \rangle_{U(N)}= \frac {{\rm e}^{\sum^Nx_\ell y_1} }
    {\ds\Delta(X)\Delta(y_2,\dots,y_N)\prod_{\ell\neq i} (x_\ell-x_i)}  
\sum_{\rho \in \mathcal
      S_{N-1}}\!\!\!\!\! \epsilon(\rho) \prod_{k=1}^{N-1} 
\int_{x_k}^{x_{k\!+\!1}}
      {\rm d}\xi_k (\xi_k-x_i) {\rm e}^{\xi_k(y_{\rho(k)\!+\!1}-y_1)}  = \\
&& =\frac {{\rm e}^{\sum^Nx_\ell y_1} }
    {\Delta(X)\Delta(y_2,\dots,y_N)\prod_{\ell\neq i} (x_\ell-x_i)}  \sum_{\rho 
\in \mathcal
      S_{N-1}} \epsilon(\rho)\\
&& \le[ \le(
\frac{x_{k\!+\!1}\!-\!x_i}{y_{\rho(k)\!+\!1}\! - \!y_1} -\frac
1{(y_{\rho(k)\!+\!1}\!-\!y_1)^2}\ri) {\rm 
e}^{x_{k\!+\!1}(y_{\rho(k)+1}\!-\!y_1)}
\!\!- \! \le(
\frac{x_{k}\!-\!x_i}{y_{\rho(k)\!+\!1}\! -\! y_1} -\frac
1{(y_{\rho(k)\!+\!1}\!-\!y_1)^2}\ri) {\rm e}^{x_{k}(y_{\rho(k)\!+\!1}\!-\!y_1)}
\ri] = \\
&& =\frac {{\rm e}^{\sum^Nx_\ell y_1} }
    {\Delta(X)\Delta(Y)\prod_{\ell\neq i} (x_\ell-x_i)\prod_{\ell\neq 1} 
(y_\ell-y_1)} 
\sum_{\rho \in \mathcal
      S_{N-1}} \epsilon(\rho)\\
&& \le[ \bigg((x_{k\!+\!1}-x_i)(y_{\rho(k)\!+\!1} - y_1) -1\bigg) {\rm 
e}^{x_{k\!+\!1}(y_{\rho(k)+1}-y_1)}
-  \bigg(
(x_{k}-x_i)(y_{\rho(k)\!+\!1} - y_1) -1\bigg) {\rm 
e}^{x_{k}(y_{\rho(k)\!+\!1}-y_1)}
\ri] =
\eea
We observe that the  following identity holds
\bea
&&\hspace{-40pt}{\rm e}^{\sum^N\!\!\!x_\ell y_1} \!\!\!\!\!\sum_{\rho \in \mathcal
      S_{N-1}}\!\!\!\!\!\!\! \epsilon(\rho)\!\!
\le[\! \bigg(\!(x_{k\!+\!1}\! -\!x_i)(y_{\rho(k)\!+\!1}\! - \!y_1)\!
  -\!1\!\bigg) {\rm e}^{x_{k\!+\!1} (y_{\rho(k)+1}\!-\!y_1)}
\!- \! \bigg(\!
(x_{k}-x_i)(y_{\rho(k)\!+\!1}\! -\! y_1) \!-\!1\!\bigg) {\rm 
e}^{x_{k}(y_{\rho(k)\!+\!1}\!-\!y_1)}
\!\ri] =\\
&& =-\det\bigg[\bigg( (x_\ell\!-\!x_i)(y_m\!-\!y_1)\!-\!1\bigg){\rm e}^{x_\ell 
y_m}\bigg]_{\ell,m=1\dots N}
\label{220}\eea
which is realized by performing elementary row operations on the
matrix inside the determinant in (\ref{220}) so that the $N\times N$ determinant 
reduces to a
$(N-1)^2$ determinant by use of Laplace's formula. \par
The more general case of the expectation value of the $(i,j)$ element 
 is obtained by permutation of the spectrum of the matrix $Y$
      so as to give the formula
\bea
\langle |U_{ji}|^2 \rangle_{U(N)}=\frac {-1}
    {\Delta(X)\Delta(Y)}
\frac{
\det\bigg[\bigg(
      (x_\ell-x_i)(y_m-y_j)-1\bigg){\rm e}^{x_\ell
        y_m}\bigg]_{\ell,m=1\dots N}}{\ds \prod_{\ell\neq i} 
(x_\ell-x_i)\prod_{\ell\neq
        j} (y_\ell-y_j)} \label{shata19}
\eea
This form of Shatashvili's formula (\ref{1.4}) for the case $n=1$ is
      remarkably simple but not suitable for our later
      purposes. Moreover it is not yet clearly equivalent to
      Morozov's formula   (\ref{moroformula}), which is what we need
      for our computation.
It will be proved in appendix \ref{appA} that the two
formul\ae\ 
are indeed equivalent.
\subsection{The correlators $\langle{\rm 
Tr}{M_1}^r{M_2}^s \rangle_{V_1,V_2}$ and their generating function}
Starting from eq. (\ref{eq2}) and using   formula (\ref{moroformula})
with $a_i = {x_i}^r$ and $b_j = {y_j}^s$ 
we obtain
\bea
&& \hspace{-1.6cm} \int_{U(N)} {\rm d}U\, \tr(X^rU^\dagger Y^s U)
 {\rm
e}^{\tr(XU^\dagger YU)}=
\sum_{i,j=1}^N {x_i}^r{y_j}^s
\int_{U(N)}{\rm d}U  |U_{ji}|^2 {\rm
e}^{\tr(XU^\dagger YU)} = \\
&& \hspace{-1.5cm}=\frac 1{\Delta(X)\Delta(Y)} 
\sum_{\rho\in S_N} \epsilon(\rho) {\rm e}^{\sum x_\ell y_{\rho(\ell)}}
\sum_{n=0}^{N-1} (-1)^{n}\!\!\!\!\!\!\!\!\!\!
\sum_{i_1<i_2<\dots<i_{n+1}}
\hspace{-20pt}
 \frac { 
\det\pmatrix{
1&\cdots&1\cr
x_{i_1}& \cdots& x_{i_{n+1}}\cr
\vdots& &\vdots\cr
x_{i_1}^{n-1}& \cdots & x_{i_{n+1}}^{n-1}\cr
x_{i_1}^r & \cdots & x_{i_{n+1}}^{r}}}{
\det\pmatrix{
1&\cdots&1\cr
x_{i_1}& \cdots& x_{i_{n+1}}\cr
\vdots& &\vdots\cr
x_{i_1}^{n}& \cdots & x_{i_{n+1}}^{n}}} 
\frac { 
\det\pmatrix{
1&\cdots&1\cr
y_{\rho(i_1)}& \cdots& y_{\rho(i_{n+1})}\cr
\vdots& &\vdots\cr
y_{\rho(i_1)}^{n-1}& \cdots & y_{\rho(i_{n+1})}^{n-1}\cr
y_{\rho(i_1)}^s & \cdots & y_{\rho(i_{n+1})}^{s}}}{
\det\pmatrix{
1&\cdots&1\cr
y_{\rho(i_1)}& \cdots& y_{\rho(i_{n+1})}\cr
\vdots& &\vdots\cr
y_{\rho(i_1)}^{n}& \cdots & y_{\rho(i_{n+1}) }^{n}}}\ ,\label{characters}
\eea
with the understandings  that we use the normalized  Haar measure over
the unitary group $U(N)$
 and that for $n=0$ the ratio of
 determinants should be $x_{i_1}^r y_{\rho(i_1)}^s$.
The first observation is that the sum over $n$ does not actually need
to be extended up to the size $N$ of the random matrices because the
determinants will vanish for $n>\min(r,s)$. The next remark is that
the ratios of determinants actually define certain totally symmetric
polynomials of their arguments of degree $r-n$ and $s-n$ respectively:
 in fact they are 
Schur polynomials corresponding to hook Young diagrams 
\be
S_r(x_{i_1},\dots,x_{i_{n+1}}) :=
\frac { 
\det\pmatrix{
1&\cdots&1\cr
x_{i_1}& \cdots& x_{i_{n+1}}\cr
\vdots& &\vdots\cr
x_{i_1}^{n-1}& \cdots & x_{i_{n+1}}^{n-1}\cr
x_{i_1}^r & \cdots & x_{i_{n+1}}^{r}}}{
\det\pmatrix{
1&\cdots&1\cr
x_{i_1}& \cdots& x_{i_{n+1}}\cr
\vdots& &\vdots\cr
x_{i_1}^{n}& \cdots & x_{i_{n+1}}^{n}}}   
= \hspace{-10pt} \sum_{a_1\leq
a_2\leq\dots\leq a_{r-n}} \prod_{k=1}^{r-n} x_{i_{a_k}} 
=\hspace{-20pt}
 \sum_{j_1+\cdots+j_{n+1}=r-n}\!\!\!\!\!\!\!\!\! \!\!\!x_{i_1}^{j_1}\cdots
 x_{i_{n+1}}^{j_{n+1}}\label{hook}\ ,
\ee
and a similar expression for the $y$ part. 
It is interesting to notice that in eq. (\ref{characters}) the
 characters of the representations of the group $GL(N)$ appear; the
 same equation could possibly be derived from the character expansion
 of the integrand. 

The formal generating function of these Schur polynomials is:
\be\label{Schurgenerating}
\sum_{r=0}^\infty {1\over x^{r+1}}S_r(x_{i_1},\dots,x_{i_{n+1}}) = \prod_{k=1}^{n+1} {1\over x-x_{i_k}}
\ee

Eq. (\ref{eq2}) now becomes

\bea
&&\int  {\rm d}\mu (M_1,M_2)
\tr(M_1^rM_2^s) = \int \prod_{i=1}^N {\rm d}\mu(x_i){\rm d}\nu(y_i)\,{\Delta(X)\Delta(Y)}
\!\!\! 
\sum_{\rho\in S_N}\!\!\! \epsilon(\rho) 
{\rm e}^{\sum x_\ell y_{\rho(\ell)}} \,\,\,\times\cr
&& \times \sum_{n=0}^{\min(r,s)} (-1)^n \!\!\!\!\!\!\!\!\!\!
\sum_{i_1<i_2<\dots<i_{n+1}} 
\!\!\!\!\!\!\!\!\!\!
S_r(x_{i_1},\dots,x_{i_{n+1}})
S_s(y_{\rho(i_1)},\dots,y_{\rho(i_{n+1})})
,
\eea
where ${\rm d}\mu(x):= {\rm e}^{-V_1(x)}{\rm d} x$ and ${\rm d}\nu(y):=
{\rm e}^{-V_2(y)}{\rm d}y$.

Using eq. (\ref{Schurgenerating}) we have:
\bea
&&\int  {\rm d}\mu (M_1,M_2)
\tr\left({1\over x-M_1}{1\over y-M_2}\right) = \cr
&&=\int \prod_{i=1}^N {\rm d}\mu(x_i){\rm d}\nu(y_i)\,{\Delta(X)\Delta(Y)}
\!\!\! 
\sum_{\rho\in S_N}\!\!\! \epsilon(\rho) 
{\rm e}^{\sum x_\ell y_{\rho(\ell)}}
\sum_{n=0}^{N-1} (-1)^n \!\!\!\!\!\!\!\!\!\!
\sum_{i_1<i_2<\dots<i_{n+1}} 
\prod_{k=1}^{n+1} {1\over (x-x_{i_k})(y-y_{\rho(i_k)})}\ .
\eea
By a relabelling of the $y$'s the sum over $\rho$ becomes an
overcounting factor $N!$:
\bea
&&\int  {\rm d}\mu (M_1,M_2)
\tr\left({1\over x-M_1}{1\over y-M_2}\right) = \cr
&&N!\int \prod_{i=1}^N {\rm d}\mu(x_i){\rm d}\nu(y_i)\,{\Delta(X)\Delta(Y)} 
{\rm e}^{\sum x_\ell y_{\ell}}
\sum_{n=0}^{N-1} (-1)^n \!\!\!\!\!\!\!\!\!\!
\sum_{i_1<i_2<\dots<i_{n+1}} 
\prod_{k=1}^{n+1} {1\over (x-x_{i_k})(y-y_{i_k})}\ .
\eea
This formula allows us to obtain the following expression for the
expectations
\bea
&&\le\langle \tr\left({1\over x-M_1}{1\over y-M_2}\right)\ri\rangle_{V_1,V_2} = \cr
&& =\frac {N!} {{\mathcal Z}_N} \int \prod_{i=1}^N {\rm d}\mu(x_i){\rm
d}\nu(y_i)\,{\Delta(X)\Delta(Y)}  
 {\rm e}^{\sum x_\ell y_{\ell}}\!\!
 \sum_{n=0}^{N-1}\!  (-1)^n
\!\!\!\!\!\!\!\!\!\!\sum_{i_1<\dots<i_{n+1}} 
\prod_{k=1}^{n+1} {1\over (x-x_{i_k})(y-y_{i_k})}
 \cr
&& ={1\over N!}
\sum_{n=0}^{N-1}(-1)^n\!\!\!\!\!\! \sum_{\sigma,\tau\in
S_N}\!\!\!\!\! \epsilon(\sigma\tau)  \!\!\!\!
\sum_{i_1<i_2<\dots<i_{n+1}} 
\int \prod_{j=1}^N \psi_{\sigma(j)}(x_j)
\phi_{\tau(j)}(y_{j}) {\rm
e}^{x_jy_{j}}
\prod_{k=1}^{n+1} {1\over (x-x_{i_k})(y-y_{i_k})}
 .\label{start}
\eea
In eq. (\ref{start}) we have used the normalized quasi-polynomial
differentials defined in 
(\ref{quasidifferentials}), the fact that, with our normalizations for
the Haar measure of $U(N)$, the partition function is 
${\mathcal Z}_N = (N!)^2\prod_{j=0}^{N-1}h_j $
(see \cite{mehta}),
and the identities
\bea
&&\Delta(X) \prod_{k=1}^N {\rm d}\mu(x_k)
=\le(\prod_{j=0}^{N-1}\sqrt{h_j}\ri) \sum_{\sigma\in S_N} 
\epsilon(\sigma)\prod_{k=1}^N \psi_{\sigma(k)}(x_k)\\
&&\Delta(Y) \prod_{k=1}^N {\rm d}\nu(y_k)  = \le(\prod_{j=0}^{N-1}\sqrt{h_j}\ri)\sum_{\tau\in S_N}
\epsilon(\tau)\prod_{k=1}^N  \phi_{\tau(k)}(y_k)\ ,
\eea
which are obtained by replacing the monomials in the Vandermonde
determinants by the biorthogonal polynomials of the same degree.
\subsubsection{Proof of formula (\ref{RScorr})}
We are now in the position of proving formula (\ref{RScorr}) in a few
strokes. Taking the coefficient of $x^ry^s$ from the formal generating
function in eq. (\ref{start}) we obtain the expression 

\bea
&&\le\langle \tr\left({M_1}^r{M_2}^s\right)\ri\rangle_{V_1,V_2} = \cr
&& ={1\over N!}
\sum_{n=0}^{\min(r,s)}(-1)^n\!\!\!\!\!\!
\sum_{i_1<i_2<\dots<i_{n+1}}  \sum_{\sigma,\tau\in
S_N}\!\!\!\!\! \epsilon(\sigma\tau) \!\!
\int \prod_{j=1}^N \psi_{\sigma(j)}(x_j)
\phi_{\tau(j)}(y_{j}) {\rm
e}^{x_jy_{j}}S_r(x_{[i]_{n+1}}) S_s(y_{[i]_{n+1}})\\
&& =\frac 1{N!}\sum_{n=0}^{\min(r,s)}(-1)^n\!\!\!\!\!\!
\sum_{i_1<i_2<\dots<i_{n+1}}  \sum_{\sigma,\tau\in
S_N}\!\!\!\!\! \epsilon(\sigma\tau) \!\!
\int \prod_{j\not \in \{i_1,\dots,i_{n+1}\}}^N \psi_{\sigma(j)}(x_j)
\phi_{\tau(j)}(y_{j}) {\rm
e}^{x_jy_{j}} \times\label{line}\\
&&\hspace{3cm}\times\int \prod_{k=1}^{n+1}\psi_{\sigma(i_k)}(x_{i_k})
\phi_{\tau(i_k)}(y_{i_k}) {\rm
e}^{x_{i_k}y_{i_k}} S_r(x_{[i]_{n+1}}) S_s(y_{[i]_{n+1}})\label{line2}
\ .
\eea
In this formula the notation $x_{[i]_{n+1}}$ means the sequence of
variables $x_{i_1},\dots,x_{i_{n+1}}$ (similarly for the $y$'s) and we
have used the fact that the Schur polynomials $S_r$ as defined in
(\ref{hook})  vanish if the
number of variables is greater than $r$.
Next, the orthogonality relations between the $\phi_n$'s and the
$\psi_n$'s in line (\ref{line})
 imply that the sum over the permutations $\sigma,\tau$ is restricted
to those permutations such that 
\be
\sigma=\tau\circ \eta\ ,\ \ \eta\in S\{i_1,\dots i_{n+1}\}
\ee
where $S\{i_1,\dots
i_{n+1}\}$ denotes the group of permutation of the indices $i_k$.
The restriction on the indices $i_1,\dots,i_{n+1}$ can be lifted
because the following expression is permutation invariant in the
label of those indices and when two such indices coincide the
corresponding term vanishes due to the alternating form of the
sum. This will produce an overcounting of a factor $(n+1)!$ which must
be corrected: moreover 
 we can relabel the variables of integration of the integral in line
 (\ref{line2})
from $x_{i_k},\ y_{i_k}$ to $x_k,\ y_k$
\bea
&&\hspace{-1cm}\le\langle \tr\left({M_1}^r{M_2}^s\right)\ri\rangle_{V_1,V_2} = \cr
&&\hspace{-1cm} =\frac 1{N!}  \sum_{\sigma\in S_N} 
\sum_{n=0}^{\min(r,s)}\frac{(-1)^n}{(n+1)!}\!\!
\sum_{i_1,i_2,\dots,i_{n+1}} \sum_{\eta\in
S_{n+1}}\!\!\!\!\! \epsilon(\eta) \!\!
\int  {\rm
e}^{\sum_k^{n+1}x_{k} y_{k}}  S_r(x_{[1,\dots]}) S_s(y_{[1,\dots]})\prod_{k=1}^{n+1} \psi_{i_k}(x_{k}) 
\phi_{i_{\eta(k)}}(y_{k})=\\
&&\hspace{-1cm}  =
\sum_{n=0}^{\min(r,s)}\frac{(-1)^n}{(n+1)!}\!\!
\sum_{i_1,i_2,\dots,i_{n+1}}  \sum_{\eta\in
S_{n+1}}\!\!\!\!\! \epsilon(\eta) \!\!\int
 {\rm
e}^{\sum_k^{n+1} x_{k} y_{k}} S_r(x_{[1,\dots, n+1]})
 S_s(y_{[1,\dots, n+1]}) \prod_{k=1}^{n+1} \psi_{i_k}(x_{k})
\phi_{i_{k}}(y_{\eta^{-1}(k)}) = \\
&&\hspace{-1cm} =
\sum_{n=0}^{\min(r,s)}\frac{(-1)^n}{(n+1)!}\!\!
\sum_{i_1,i_2,\dots,i_{n+1}}  \sum_{\eta\in
S_{n+1}}\!\!\!\!\! \epsilon(\eta) \!\!
\int  {\rm
e}^{\sum_k^{n+1}x_{k} y_{k}}  S_r(x_{[1,\dots, n+1]}) S_s(y_{[1,\dots,
    n+1]})\prod_{k=1}^{n+1} \psi_{i_k}(x_{k}) 
\phi_{i_{\eta(k)}}(y_{k})=\\
&&\hspace{-1cm}  =
\sum_{n=0}^{\min(r,s)}\frac{(-1)^n}{(n+1)!}\!\!
\sum_{i_1,i_2,\dots,i_{n+1}}  \sum_{\eta\in
S_{n+1}}\!\!\!\!\! \epsilon(\eta) \!\!\int
 {\rm
e}^{\sum_k^{n+1} x_{k} y_{k}} S_r(x_{[1,\dots, n+1]})
 S_s(y_{[1,\dots, n+1]}) \prod_{k=1}^{n+1} \psi_{i_k}(x_{k})
\phi_{i_{k}}(y_{\eta^{-1}(k)}) = \\
&& = \sum_{n=0}^{\min(r,s)}\frac{(-1)^n}{(n+1)!}\!\!
\int
 {\rm
e}^{\sum_k^{n+1} x_{k} y_{k}} S_r(x_{[1,\dots, n+1]})
 S_s(y_{[1,\dots, n+1]})
 \det\bigg[K_{12}(x_j,y_\ell)\bigg]_{j,\ell\leq n+1}
\eea
where we have used the
definition of the kernel $K_{12}(x,y)$ given in
eq. (\ref{kernel}).
This concludes the proof of eq. (\ref{RScorr}) from which formula
(\ref{fgcorr}) follows immediately.
\subsubsection{Proof of formula (\ref{beauty})}
Resuming from eq. (\ref{start}) and performing 
a relabelling of the $x$'s and the $y$'s allows to choose $i_k=k$, and the
sum over $i_{1}<\dots<i_{n+1}$ becomes a combinatorial factor:
\bea
&&\le\langle \tr\left({1\over x-M_1}{1\over y-M_2}\right)\ri\rangle_{V_1,V_2} =
\cr&& ={1\over N!}
\sum_{n=0}^{\infty}(-1)^n \pmatrix{N\cr n+1} \!\! \sum_{\sigma,\tau\in
S_N}\!\!\!\!\! \epsilon(\sigma\tau) 
\int \prod_{j=1}^N \psi_{\sigma(j)}(x_j)
\phi_{\tau(j)}(y_{j}) {\rm
e}^{x_jy_{j}}
\prod_{k=1}^{n+1} {1\over (x-x_{k})(y-y_{k})}
.
\eea
Now, the equations (\ref{norms}) and (\ref{mulxy}) imply that
\be
\int  \psi_{\sigma(j)}(x')
\phi_{\tau(j)}(y') {\rm e}^{x'y'}
= \delta_{\sigma(j),\tau(j)} \, ,
\ee 
and
\be
\int  \psi_{\sigma(j)}(x')
\phi_{\tau(j)}(y') {\rm e}^{x'y'}
{1\over (x-x')(y-y')}
= W_{\sigma(j),\tau(j)}
\ee
where $W$ is the $N\times N$ square matrix\footnote{We remind the
  reader that this matrix should be properly understood as a formal
  power series in inverse powers of $x$ and $y$, although an analytic
  definition could be given in terms of the biorthogonal
  polynomials. However this is unnecessary for the scope of the
  present paper.}:
\be
W:={\ds \m{\pi}_N}{1\over x-Q}{1\over y-P}{\ds \m{\pi}_N}^t\ .
\ee
Therefore $\tau=\sigma\eta$ where $\eta$ is a permutation of the $n+1$ first indices
only, and we can write:
\bea
\le\langle \tr\left({1\over x-M_1}{1\over y-M_2}\right)\ri\rangle_{V_1,V_2} 
 & = &{1\over N!}
\sum_{n=0}^{\infty}(-1)^n \pmatrix{N\cr n+1} \!\! \sum_{\sigma\in S_N}
\sum_{\eta\in S_{n+1}}
\!\! \epsilon(\eta) 
\prod_{j=1}^{n+1} W_{\sigma(j),\sigma\eta(j)}
\cr
& = &
\sum_{n=0}^{\infty}{(-1)^n \over( n+1)! (N-n-1)!} \!\! \sum_{\sigma\in S_N}
\det \left(W_{\sigma(i),\sigma(j)}\right)_{i,j=1,\dots,n+1}
.
\eea
We note $\alpha_i=\sigma(i)$
and we write:
\be
\le\langle \tr\left({1\over x-M_1}{1\over y-M_2}\right)\ri\rangle_{V_1,V_2} 
= 
\sum_{n=0}^{\infty}{(-1)^n \over (n+1)! (N-n-1)!} \!\! 
\sum_{\alpha_1\neq\dots\neq\alpha_N}
\det \left(W_{\alpha_i,\alpha_j}\right)_{i,j=1,\dots,n+1}
,
\ee
where the sum over $\alpha_{n+2},\dots,\alpha_{N}$ disappears and brings a factor $(N-n-1)!$:
\bea
\le\langle \tr\left({1\over x-M_1}{1\over y-M_2}\right)\ri\rangle_{V_1,V_2} 
& = & 
\sum_{n=0}^{\infty}{(-1)^n \over (n+1)! } \!\! 
\sum_{\alpha_1\neq\dots\neq\alpha_{n+1}}
\det \left(W_{\alpha_i,\alpha_j}\right)_{i,j=1,\dots,n+1} \cr
& = & 
\sum_{n=0}^{\infty}{(-1)^n \over( n+1)! } \!\! 
\sum_{\alpha_1\neq\dots\neq\alpha_{n+1}}
\sum_{\eta\in S_{n+1}} \epsilon(\eta)
\prod_{j=1}^{n+1}
W_{\alpha_i,\alpha_{\eta(j)}}
,
\eea
and we can replace the sum over distinct $\alpha$'s by a sum over all $\alpha$'s:
\be
\le\langle \tr\left({1\over x-M_1}{1\over y-M_2}\right)\ri\rangle_{V_1,V_2} 
= 
\sum_{n=0}^{\infty}{(-1)^n \over (n+1)! } \!\! 
\sum_{\eta\in S_{n+1}} \epsilon(\eta)
\sum_{\alpha_1,\dots,\alpha_{n+1}}
\prod_{j=1}^{n+1}
W_{\alpha_i,\alpha_{\eta(j)}}
.
\ee
If $\eta$ is decomposed into a product of ${\cal N}(\eta)$ cyclic permutations
of lengths
$l_1(\eta)+\dots+l_{{\cal N}(\eta)}(\eta)=n+1$, we have:
\bea
\le\langle \tr\left({1\over x-M_1}{1\over y-M_2}\right)\ri\rangle_{V_1,V_2} 
& = &
\sum_{n=0}^{\infty}{(-1)^n \over (n+1)! } \!\! 
\sum_{\eta\in S_{n+1}} 
\prod_{j=1}^{{\cal N}(\eta)} (-1)^{l_j(\eta)+1}
\tr W^{l_j(\eta)}\cr
& = &
-\sum_{n=0}^{\infty}{ 1\over (n+1)! } \!\! 
\sum_{\eta\in S_{n+1}} \prod_{j=1}^{{\cal N}(\eta)} (-\tr W^{l_j(\eta)})
,
\eea

Now we use the following Lemma, which is a classical result in combinatorics:

\bl
Let $\mathcal G_m$ be a function defined on the permutation group of
$m$ elements $S_m$ with the {\em cluster property}, i.e., such that if 
$\eta=\eta_1\circ \eta_2$
is a decomposition into disjoint permutations of $m'$ and $m''$
elements (and hence $m=m'+m''$) then
\be
\mathcal G_m(\eta) = \mathcal G_{m'}(\eta_1) \mathcal G_{m''}(\eta_2)\ .
\ee    
Under these circumstances  we have the identity
\be
\exp\le(\sum_{m=1}^{\infty} \sum_{\sigma \in \mathcal C_m} \frac
{x^m}{m!} \mathcal G_m(\sigma)\ri) = 1+ \sum_{m=1}^{\infty}\frac 
{x^{m}}{m!} \sum_{\eta\in S_m} \mathcal G_m(\eta)\ ,\label{cluster}
\ee
where $\mathcal C_m$ denotes the set of all permutations of maximal
length and has cardinality $(m-1)!$.
\label{clusterlemma}
\el
In other words this lemma says that if $\mathcal G$ has 
the cluster property, then taking the logarithm of the RHS of eq. 
(\ref{cluster}) removes 
all ``nonconnected'' contributions and returns the ``connected
components'' only. 
In view of Lemma \ref{clusterlemma}  let us define
\be
\mathcal G_m(\eta) := \prod_{j=1}^{{\cal N}(\eta)} (-\tr W^{l_j(\eta)})
\ee
which has clearly the cluster property, and we have:
\bea
&& 1-\le\langle \tr\left({1\over x-M_1}{1\over y-M_2}\right)\ri\rangle_{V_1,V_2}
 =   1+ \sum_{m=1}^{\infty}\frac 
{1}{m!} \sum_{\eta\in S_m} \mathcal G_m(\eta) 
 =  \exp{\left[ \sum_{m=1}^{\infty} \sum_{\eta \in \mathcal C_m} \frac
{1}{m!} \mathcal G_m(\eta) \right]} \cr
&& =  \exp{\left[ -\sum_{m=1}^{\infty}  \frac{\tr W^m}{m}  \right]} 
 =  \exp{\left[ \tr \ln{(\1_{N}-W)}  \right]} 
 =  \det (\1_{N}-W) 
\eea
This concludes our proof of formula (\ref{beauty}).
\section{Conclusions}
Formula (\ref{beauty}) is quite simple in spite of the long
computations involved in its proof. It is tempting to imagine that
also more complicated multi-correlators could be reduced to a
computation involving the matrices $P$ and $Q$, i.e., to biorthogonal
polynomials. Quite clearly the possibility rests on having a
manageable formula for the generic multi-correlators of the
Itzykson-Zuber-Harish-Chandra integral. There are indications that such
a formula should be derivable: for example it is quite
simple to obtain a  formula for a correlator of entries with the same
first index. Indeed starting from eq. (\ref{1.4}) and proceeding in
the same way we did in order to obtain eq. (\ref{shata19}) one can
easily prove the following formula
\bea
\langle\prod_{a=1}^n |U_{j\,i_a}|^2 \rangle_{U(N)} = 
\frac {(-1)^n}{n!}\frac { \det\bigg[F_m(x_\ell){\rm e}^{x_\ell
    y_m}\bigg]_{\ell,m=1\dots N}
}
    {\ds\Delta(X)\Delta(Y)\prod_{a=1}^n \prod_{\ell\neq i_a}
      (x_\ell-x_{i_a})\prod_{\ell\neq j}(y_\ell-y_j)^n}
\eea
where 
\bea
F_{m}(\xi):=\bigg[ (y_{m}\!-\!y_j)^n \prod_{a=1}^n(\xi\!-\!x_{i_a}) \! -\!
  (y_{m}\!-\!y_j)^{n-1}\sum_{\ell=1}^n\prod_{a\neq\ell}^n(\xi\!-\!x_{i_a})\! +\!\dots \!+\!(-1)^n n!
\bigg]  =\\
=\sum_{s=0}^n (-1)^s (y_{m}\!-\!y_j)^{n-s}\frac {{\rm d}^s}{{\rm
     d}\xi^s} \prod_{a=1}^n(\xi\!-\!x_{i_a})\ .
\eea
However the computation for non-equal first indices becomes quickly
extremely complicated at least using the technique in \cite{shata}.
Nonetheless we hope that this first computation can break through the
general belief that computations of correlators in multi-matrix models
are not feasible in the finite-N regime due to the angular
integrations.\par
It should also be remarked that for polynomial potentials we could use
the so-called ``loop'' equations to obtain information on other
correlators. The result, however would be dependent on the specific
form of the potentials and would not provide information on the HCIZ
integral itself.

\medskip

Let us also mention that this calculation could be generalized to other random
 matrix models, in particular the complex matrix model, which presents many
 similarities with the 2-matrix model.
 The gaussian complex matrix model has attracted lot of attention in string
  theory.
A particular case of the ADS/CFT, is the conjectured duality between 
string-theory in a pp-wave background, and BMN gauge theory.
The gaussian complex-matrix model appears as an effective BMN theory 
in a particular limit, and the computation of mixed correlation functions is very important
 in that model \cite{Berenstein, eynardkris}. 
In the gaussian complex matrix model, a formula is known for the 2-point mixed correlator
$\langle \tr{1\over x-M}{1\over y-M^\dagger}\rangle$, but little is known
for other mixed correlation functions.

\appendix
\section{Equivalence of eq. (\ref{moroformula}) and (\ref{shata19})}
\label{appA}
%%%%%%%%%%%%%%%%%%%%%%%%%%%%%%%%%%%%%%%%%
%%%%%%%%%%%%%%%%%%%%%%%%%%%%%%%%%%%%%%%%%
%%%%%%%%%%%%%%%%%%%%%%%%%%%%%%%%%%%%%%%%%
%%%%%%%%%%%%%%%%%%%%%%%%%%%%%%%%%%%%%%%%%
%%%%%%%%%%%%%%%%%%%%%%%%%%%%%%%%%%%%%%%%%
%%%%%%%%%%%%%%%%%%%%%%%%%%%%%%%%%%%%%%%%%

 In this appendix we prove that the two formul\ae\ (\ref{moroformula})
      and (\ref{shata19}) are equivalent.
To this end we start from (\ref{shata19}) and  compute
\bea
\langle |U_{ji}|^2\rangle_{U(N)}  = \frac {-1}
    {\Delta(X)\Delta(Y)}
\frac{
\det\bigg[\bigg(
      (x_\ell-x_i)(y_m-y_j)-1\bigg){\rm e}^{x_\ell
        y_m}\bigg]_{\ell,m=1\dots N}}{\ds \prod_{\ell\neq i} 
(x_\ell-x_i)\prod_{\ell\neq
        j} (y_\ell-y_j)}=\\
=\frac {-1}
    {\Delta(X)\Delta(Y)}
 \sum_{\rho\in S_{N}} \epsilon(\rho)
   \frac{\ds {\rm e}^{\sum x_\ell y_{\rho(\ell)}}} {\ds \prod_{\ell\neq
        i} (x_\ell-x_i)\prod_{\ell\neq
        j} (y_\ell-y_j)}  \prod_{\ell=1}^{N} \bigg(
      (x_\ell-x_i)(y_{\rho(\ell)}-y_j)-1\bigg)=\\
=\frac {-1}
    {\Delta(X)\Delta(Y)}
 \sum_{\rho\in S_{N}} \epsilon(\rho)
 \frac{\ds (-1)^N {\rm e}^{\sum x_\ell y_{\rho(\ell)}}} {\ds \prod_{\ell\neq
        i} (x_\ell-x_i)\prod_{\ell\neq
        j} (y_\ell-y_j)}   \prod_{\ell=1}^{N} \bigg(1-
      (x_\ell-x_i)(y_{\rho(\ell)}-y_j)\bigg)=\\
=\frac {1}
    {\Delta(X)\Delta(Y)}
 \sum_{\rho\in S_{N}} \epsilon(\rho)
 \frac{\ds (-1)^{N+1} {\rm e}^{\sum x_\ell y_{\rho(\ell)}}} {\ds \prod_{\ell\neq i} 
(x_\ell-x_i)\prod_{\ell\neq j} (y_\ell-y_j)} \\
\Bigg[1 - \sum_{\ell=1}^N  (x_\ell-x_i)(y_{\rho(\ell)}-y_j) +
  \sum_{\ell_1 <\ell_2}  (x_{\ell_1}-x_i)(y_{\rho(\ell_1)}-y_j)
 (x_{\ell_2}-x_i)(y_{\rho(\ell_2)}-y_j) + \dots \Bigg]=\\
\frac {1}
    {\Delta(X)\Delta(Y)}
 \sum_{\rho\in S_{N}} \epsilon(\rho)
 \frac{\ds (-1)^{N+1} {\rm e}^{\sum x_\ell y_{\rho(\ell)}}} {\ds \prod_{\ell\neq i} 
(x_\ell-x_i)\prod_{\ell\neq j} (y_\ell-y_j)} 
\Bigg[1 + \sum_{n=1}^N (-1)^{n}\!\!\!\!\!\!\!\!\!\!\! 
\sum_{\ell_1<\ell_2<\dots<\ell_n}
  \prod_{k=1}^n  (x_{\ell_k}-x_i)(y_{\rho(\ell_k)}-y_j)
  \Bigg]= \\
\frac {1}
    {\Delta(X)\Delta(Y)}
 \sum_{\rho\in S_{N}} \epsilon(\rho)
 \frac{\ds (-1)^{N+1} {\rm e}^{\sum x_\ell y_{\rho(\ell)}}} {\ds \prod_{\ell\neq i} 
(x_\ell-x_i)\prod_{\ell\neq j} (y_\ell-y_j)} 
\Bigg[1 + \sum_{n=1}^N (-1)^{n}\!\!\!\!\!\!\!\!\!\!\! 
\sum_{\ell_1<\ell_2<\dots<\ell_n \atop
    \ell_k\neq i, \rho(\ell_k)\neq j,\forall k}
  \prod_{k=1}^n  (x_{\ell_k}-x_i)(y_{\rho(\ell_k)}-y_j)
  \Bigg]\label{shata27}
\eea 
Let us consider the following subexpression from the above formula
\bea
 \hspace{-1truecm} \frac {(-1)^{N+1}} {\ds \prod_{\ell\neq i} 
(x_\ell-x_i)\prod_{\ell\neq j} (y_\ell-y_j)} 
\Bigg[1 + \sum_{n=1}^N (-1)^{n}\!\!\!\!\!\! \sum_{\ell_1<\ell_2<\dots<\ell_n \atop
    \ell_k\neq i, \rho(\ell_k)\neq j,\forall k}
  \prod_{k=1}^n  (x_{\ell_k}-x_i)(y_{\rho(\ell_k)}-y_j)\Bigg] 
\stackrel{(\star)}{=} \\
 \hspace{-1truecm}=(-1)^{N+1}
\le[ \frac 1 {\ds \prod_{\ell\neq i} 
(x_\ell-x_i)\prod_{\ell\neq j} (y_\ell-y_j)}  + \sum_{n=1}^{N-1}
  (-1)^n \!\!\!\!\!\!\!\sum_{\ell_1<\ell_2<\dots<\ell_n \atop 
    \ell_k\neq i, \rho(\ell_k)\neq j,\forall k}\frac{\ds
  \prod_{k=1}^n  (x_{\ell_k}-x_i)(y_{\rho(\ell_k)}-y_j)} {\ds \prod_{\ell\neq i} 
(x_\ell-x_i)\prod_{\ell\neq j} (y_{\rho(\ell)}-y_j)}   \ri]  = \\
=(-1)^{N+1}\!\!
\le[  \frac 1 {\ds \prod_{\ell\neq i} 
(x_\ell-x_i)\prod_{\ell\neq j} (y_\ell-y_j)}\! +\!\!\!\! \sum_{n=1}^{N-1}
(-1)^n\!\!\!\!\!\!\!\!\!\!\!\!
 \sum_{\ell_1<\ell_2<\dots<\ell_n \atop
    \ell_k\neq i, \rho(\ell_k)\neq j,\forall k}\frac 1 {\ds
    \prod_{\ell\not\in \{i,\ell_k,\forall k\}} 
(x_\ell-x_i)\!\!\!\!\!\!\!\prod_{\rho(\ell)\not \in\{ j,\rho(\ell_k),\forall
  k\}} \!\!\!\!\!\!(y_{\rho(\ell)}-y_j)}   \ri]  = 
\eea
\bea
=\le[  \frac {(-1)^{N+1}} {\ds \prod_{\ell\neq i} 
(x_\ell-x_i)\prod_{\ell\neq j} (y_\ell-y_j)} + \sum_{n=1}^{N-1} 
(-1)^{N-n+1}\!\!\!\!\!\!
\sum_{i_1<\dots<i_{N-n}: \atop i\in \{i_k\},
  j\in\{\rho(i_k)\} }\frac 1 {\ds
    \prod_{k: i_k\neq i} 
(x_{i_k}-x_i)\prod_{k:\rho(i_k)\neq j}  (y_{\rho(i_k)}-y_j)}   \ri]
= \\
=
\le[  \frac {(-1)^{N+1}} {\ds \prod_{\ell\neq i} 
(x_\ell-x_i)\prod_{\ell\neq j} (y_\ell-y_j)} + \sum_{n=1}^{N-1} 
(-1)^{n+1}\!\!\!\!\!\!\!
\sum_{i_1<\dots<i_{n}: \atop i\in \{i_k\},
  j\in\{\rho(i_k)\} }\frac 1 {\ds
    \prod_{k: i_k\neq i} 
(x_{i_k}-x_i)\prod_{k:\rho(i_k)\neq j}  (y_{\rho(i_k)}-y_j)}
\ri]=\\
=
\le[  \sum_{n=1}^{N} (-1)^{n+1}\!\!\!\!\!\!\!
\sum_{i_1<\dots<i_{n}: \atop i\in \{i_k\},
  j\in\{\rho(i_k)\} }\frac 1 {\ds
    \prod_{k: i_k\neq i} 
(x_{i_k}-x_i)\prod_{k:\rho(i_k)\neq j}  (y_{\rho(i_k)}-y_j)}
\ri]=\\
\stackrel{(\star\star)}{=}
\le[  \sum_{n=1}^{N} (-1)^{n+1}\!\!\!\!\!\!\!
\sum_{i_1<\dots<i_{n} }\!\!\!\!\!\!\frac {\ds\sum_{s=1}^n (-1)^s \delta_{i,i_s}
  \Delta(x_{i_1},\dots,\widehat {x_{i_s}},\dots x_{i_n})\sum_{s=1}^n (-1)^s 
\delta_{j,\rho(i_s)}
  \Delta(y_{\rho(i_1)},\dots,\widehat {y_{\rho(i_s)}},\dots y_{\rho(i_n)})}  
{\ds
   \Delta(x_{i_1},\dots,x_{i_n}) \Delta(y_{\rho(i_1)},\dots,y_{\rho(i_n)})}
\ri]=\\
\hspace{-1cm}=  \sum_{n=1}^{N} (-1)^{n+1}\!\!\!\!\!\!\!
\sum_{i_1<\dots<i_{n} }
 \frac { 
\det\pmatrix{
1&\cdots&1\cr
x_{i_1}& \cdots& x_{i_{n+1}}\cr
\vdots& &\vdots\cr
x_{i_1}^{n-1}& \cdots & x_{i_{n+1}}^{n-1}\cr
\delta_{ii_1} & \cdots & \delta_{ii_{n+1}}}}{
\det\pmatrix{
1&\cdots&1\cr
x_{i_1}& \cdots& x_{i_{n+1}}\cr
\vdots& &\vdots\cr
x_{i_1}^{n}& \cdots & x_{i_{n+1}}^{n}}} 
\frac { 
\det\pmatrix{
1&\cdots&1\cr
y_{\rho(i_1)}& \cdots& y_{\rho(i_{n+1})}\cr
\vdots& &\vdots\cr
y_{\rho(i_1)}^{n-1}& \cdots & y_{\rho(i_{n+1})}^{n-1}\cr
\delta_{j\rho(i_1)} & \cdots & \delta_{j\rho(i_{n+1})}}}{
\det\pmatrix{
1&\cdots&1\cr
y_{\rho(i_1)}& \cdots& y_{\rho(i_{n+1})}\cr
\vdots& &\vdots\cr
y_{\rho(i_1)}^{n}& \cdots & y_{\rho(i_{n+1}) }^{n}}}\label{final1}
\eea
In the above chain of equality we have replaced (after the $(\star)$)
the upper limit of summation by $N-1$ because in the case $n=N$ there
is certainly one $\ell_k$ equal to $i$ (and one $\rho(\ell_k)$ equal
to $j$) so that that term does not contribute. After $(\star\star)$ we
have removed the condition on the multi-index because it is implicit
in the sum of deltas that follows.
Putting together eq. (\ref{final1}) with eq. (\ref{shata27}) or
(\ref{shata19})  we obtain
the desired proof of the equivalence of formula (\ref{moroformula})
with eq. (\ref{1.4}), thus also proving the former rigorously, which
was not done in \cite{morozov}.

\end{document}